\documentstyle[11pt]{article}

\newcommand{\beq}{\begin{equation}}
\newcommand{\eeq}{\end{equation}}
\newcommand{\beqa}{\begin{eqnarray}}
\newcommand{\eeqa}{\end{eqnarray}}

\title{Naturally Light Scalars}
\author{John Terning\\
Department of Physics, Boston University\\
590 Commonwealth Ave., Boston MA  02215
}

\begin{document}
\setlength{\baselineskip}{24pt}
\maketitle
\begin{picture}(0,0)(0,0)
\put(295,239){BUHEP-95-28}
\end{picture}
\vspace{-12pt}
\date{}
\begin{abstract}I argue that in certain chiral gauge theories
composite scalars associated with chiral symmetry breaking
can be light (i.e. lighter than
naive scaling from QCD would suggest) without any fine-tuning.
These scalars will be even lighter in chiral gauge
theories that produce chiral symmetry breaking without confinement.
I construct a model which demonstrates this last possibility.
\end{abstract}

\section{Introduction}
The fact that scalar mesons are much heavier than pseudoscalar and
vector mesons was once considered to be something of a mystery.
In fact,
a mythical $\epsilon(730)$ once graced the pages of the Particle Data
Book \cite{oldPDG}, in conformity with certain theoretical prejudices.
Currently the lightest scalar composed primarily of a quark and
anti-quark
is thought to weigh in between 1300 and 1500 MeV \cite{PDG}.
(The $f_0(980)$ and   $a_0(980)$,
formerly known as the $S^*$ and $\delta$, are thought to be primarily
$K$-${\overline K}$ bound states \cite{KK,Hill}.)
Recently Hill and Marinelli \cite{Hill} pointed out that in QCD
certain scalar mesons (those which
are the parity partners of the pseudo-Goldstone bosons)
receive a contribution to their masses from instantons, and thus,
like the $\eta^\prime$, their masses can be larger than  might be
naively expected.  In this paper I will explore the
implications of Hill and Marinelli's
observation for the spectra of certain chiral gauge theories. I will
also consider chiral gauge theories that undergo chiral symmetry breaking
without confinement.  I will show that scalars can be quite light in certain
chiral gauge theories.  Such light scalars may be of use in building
extensions of the standard model.

\section {The Effect of Instantons}
As noted by Hill and Marinelli \cite{Hill}, the effective interaction induced
on quarks by instantons (the t'Hooft determinant interaction
\cite{tHooft})
not only contributes to the
$\eta^\prime$ mass, but also to the scalar meson masses.  I will
roughly estimate the size of this contribution to the $\eta^\prime$
mass
by assuming that in the absence of instantons the $\eta^\prime$
is a pseudo-Goldstone boson. Neglecting the up and down quark masses,
the mass (squared) matrix for the $\eta$ and $\eta^\prime$ is:
\beq
M^2 = {{< {\overline \psi}\psi>}\over
{3 f^2}}  m_s \left( \begin{array}{cc}
4 & -2\sqrt{2} \\
 -2\sqrt{2} & 2 \end{array}
\right)+ {1\over 2}(I-\tau_3)M_I^2 ~,
\eeq
where the first term is a standard chiral perturbation theory
estimate \cite{Wilczek}, $\tau_3$ is a Pauli matrix,
and $M_I^2$ represents the instanton contribution.

Writing
\beq
M_I^2 =  a \, {{< {\overline \psi}\psi>}\over
{ f^2}}  m_s~,
\eeq
I find
\beq
m_{\eta^\prime}^2 + m_{\eta}^2 = (2 +a) \,m_s {{< {\overline \psi}\psi>}\over
{ f^2}} ~,
\label{sum}
\eeq
\beq
m_{\eta^\prime}^2 - m_{\eta}^2=m_s {{< {\overline \psi}\psi>}\over
{ f^2}}  \sqrt{4-{4\over 3}a + a^2} ~.
\label{diff}
\eeq
Using standard values for strange quark mass and the condensate
($m_s=155 \, MeV$, $<{\overline \psi}\psi>=(236 \, {\rm MeV})^3$),
equation (\ref{sum}) gives $M_I \approx 870$ MeV ($a = 3.16$); while
this result verifies equation (\ref{diff}) to within 20\%.
If the contribution to the scalar mass (squared) is comparable to $M_I^2$,
then in the
absence of instantons, the scalar mass would be reduced to the range
970-1220 MeV.

\section {Suppressing Instantons and the Effect of Confinement}
How would things be different in chiral gauge theories?  Recall
that the effective interaction induced by instantons takes the
form of a determinant, involving each of the (left and right-handed)
fermions with non-Abelian gauge interactions.  In producing a mass
term in an effective Lagrangian for mesons, four of the fermions are
replaced by two meson
fields, and the remaining fermion lines must be contracted.  This is
only possible if all of the remaining fermions get Dirac masses (i.e.
the condensate  $<\overline{\psi_L} \psi_R>$ is non-zero).  But
chiral gauge theories are characterized by the fact that not all the
fermions can have gauge invariant mass terms.  The generic situation
for chiral symmetry breaking in chiral gauge theories
is that not all the fermions get masses at the same scale. Thus,
if the chiral gauge theory has more than four interacting
fermions (counting left and right-handed fermions separately), and
not all the fermions get dynamical masses\footnote{A special case of
this situation is when there is an odd number of fermions.} then at
least some of the
scalars will not get a contribution to their masses from instantons.
If the fermions that are massless at the chiral symmetry
breaking scale under consideration ($\Lambda_h$ the ``heavy" scale)
actually develop a mass at some
lower scale ($\Lambda_l$ the ``light" scale), then the instanton
contribution to the
scalar mass squared will be
suppressed\footnote{This suppression also applies to the mass squared of the
analogue of the $\eta^\prime$.} by a power of the
light scale over the heavy scale.  That is:
\beq
M_I^2 = \kappa \, \Lambda_h^2
\left({{\Lambda_l}\over{\Lambda_h}}\right)^{3n}
{}~,
\eeq
where $n$ is just the number of light Dirac fermions.

As shown above, even in the absence of instanton contributions to their
masses, the scalars can still be quite
heavy, compared to the vector mesons.
To proceed further, we must understand more about what makes scalars
heavy.  The heuristic explanation is that confinement makes them
heavy.  Although no one understands why non-relativistic quark models
are so successful in describing the spectrum of hadrons containing light
quarks, it will be helpful to consider what such models have to say about
this issue.  From the viewpoint of non-relativistic potential models
the answer is quite clear. In a Coulombic potential the $1S$ (i.e. the
analogues of the $\rho$ and $\pi$) has
a small splitting (${\cal O}(\alpha^2)$) from the degenerate $2S$ and
$2P$ (i.e. the analogues of the $\omega$ and the scalar) levels. However, in a
confining potential (e.g. a linear or even
harmonic potential) the $1S$ can have a large splitting from the $2P$ (which is
not degenerate with the $2S$).

Whether or not the results of non-relativistic quark models are
trustworthy for the spectra of generic chiral gauge theories, one thing is
certain: if the gauge symmetry left unbroken by the chiral symmetry
breaking is not asymptotically free, then the long-distance contribution
to the scalar mass can be calculable (since the coupling can be weak).
In the case of a weak long-range coupling, a Coulombic potential should
be a good approximation.  In the next section I will construct a chiral
gauge theory that, according to conventional wisdom, should
demonstrate this behavior: it produces chiral symmetry
breaking without confinement.  Or in other words, it is asymptotically
free above the chiral symmetry breaking scale, but not asymptotically
free below this scale.

\section{A Toy Model}
In this section I will construct a model that dynamically breaks chirally
symmetries, without
confinement.
Let ${\bf S_N}$ be the symmetric tensor representation of $SU(N)$, then
the following  reducible representations:
\beq
{\rm(N+4)} \times {\bf N} \oplus {\bf \overline{S_N}}~,
\label{redrep}
\eeq
and
\beq
 {\bf N} \oplus {\bf \overline{N}},
\label{vecrep}
\eeq
are each anomaly free.
According to the most attractive channel (MAC) hypothesis \cite{MAC},
a condensate  will form first in the MAC as
determined by the strength of one gauge boson exchange\footnote{Of course, the
MAC hypothesis could be wrong,  see ref. \cite{chiralTC} for a recent
critique.}, i.e. the sum of the
quadratic Casimirs of the fermions minus the quadratic Casimir of the
condensate, which I will call $\Delta C_2$.
The strength of the possible condensation channels for the
fermions discussed above are:
\beqa
\Delta C_2({\bf N} \times {\bf \overline{S_N}} \rightarrow {\bf
\overline{N}})&=&{{(N+2)(N-1)}\over{N}}~,\\
\Delta C_2({\bf N} \times {\bf \overline{N}} \rightarrow {\bf
1})&=&{{(N+1)(N-1)}\over{N}}~,\\
\Delta C_2({\bf \overline{S_N}} \times {\bf \overline{S_N}} \rightarrow {\bf
\overline{R_1}})&=&{{2(N+2)}\over{N}}~,\\
\Delta C_2({\bf \overline{N}} \times {\bf \overline{S_N}} \rightarrow {\bf
\overline{R_2}})&=&{{(N+3)}\over{N}}~,\\
\Delta C_2({\bf N} \times {\bf N} \rightarrow {\bf
A})&=&{{N+1}\over{N}}~,
\label{deltacs}
\eeqa
where ${\bf A}$ is the antisymmetric tensor representation,
${\bf R_1}$ is the representation with two symmetric and two antisymmetric
indices,
i.e. its Young tableaux is composed of two rows of two boxes, and ${\bf R_2}$
has a Young
tableaux with two boxes in the first row, and one box in the second
row.  For $N>3$, the MAC is
\beq
{\bf N} \times {\bf \overline{S_N}} \rightarrow {\bf \overline{N}}
\label{cond}
\eeq
which, for $N >2$,
breaks $SU(N)$ to $SU(N-1)$.
Under this pattern of gauge symmetry breaking,
${\bf \overline{S_N}}$ decomposes into an ${\bf \overline{N-1}}$,
an ${\bf \overline{S_{N-1}}}$, and a ${\bf 1}$.
Thus the condensation in  (\ref{cond}) leaves (for each of the
reducible, anomaly-free representations shown in (\ref{redrep}))
the following  massless fermions:
$(N+5)$ ${\bf 1}$'s, $(N+3)$ ${\bf N-1}$'s,
and one  ${\bf \overline{S_{N-1}}}$.

With $n_\chi$ copies of the chiral, reducible, anomaly-free representation
shown in (\ref{redrep}),
and $n_f$ of the vector representations  (\ref{vecrep}),
the coefficient of
the one-loop $\beta$ function for $SU(N)$ is proportional to
\beq
11 N - n_\chi\left((N+4) + (N+2)\right) -2 n_f
\eeq
 For the unbroken $SU(N-1)$, below the condensation scale, the
one-loop $\beta$ function coefficient is proportional to:
\beq
11 (N-1) - n_\chi\left((N-1+4) + (N-1+2)\right)-2 n_f
\eeq
 For the  $SU(N)$ gauge group to be asymptotically free while the $SU(N-1)$
subgroup is not
(i.e.  the $SU(N-1)$ subgroup is infrared free)
the following  inequality  must be satisfied (taking $n_\chi=1$ for simplicity)
\beq
{{9N-15}\over{2}} < n_f < {{9N-6}\over{2}}~,
\eeq
which has solutions for arbitrary $N$. Thus it is very simple to arrange for
the properties
I want at one loop.  However, one must be more careful, since
for certain values of $N$, $n_\chi$, and $n_f$, there can be an infrared fixed
point in the
two-loop $\beta$ function
for small values of the gauge coupling
$\alpha = g^2/(4 \pi)$. For example, in the large $N$ limit,
with $n_\chi=1$ and $n_f = 1+(9N-15)/2$, the fixed point occurs at
\beq
\alpha_* \approx {{28 \pi}\over{39 N^2}}~.
\eeq
Thus at two loops, this theory is asymptotically free for $\alpha < \alpha_*$,
and  not asymptotically free for $\alpha > \alpha_*$.  For comparison,
the standard crude estimate (calculated in the ladder ``approximation")
of the strength of the gauge coupling
required for chiral symmetry breaking\footnote{That is the strength of
the gauge coupling required to make the anomalous dimension of
$\overline{\psi} \psi$ equal to one \cite{Rainbow}.} is:
\beq
\alpha_c(N)={{2 \pi} \over {3 \,  \Delta C_2 }}= {{2 \pi N}\over{3 (N+2)(N-1)}}
 ~.
\label{critcoup}
\eeq
Thus in the large $N$ limit with $n_\chi=1$ and $n_f=1+(9N-15)/2$, the coupling
approaches its infrared fixed point from below, and this fixed point  is
at a value too weak for chiral symmetry breaking to occur, so the low-energy
effective theory is a (massless) conformal theory.

To produce a model that
exhibits chiral symmetry breaking without confinement, one must choose $N$,
 $n_\chi$, and
$n_f$ such that
\beq
 \alpha_*(N,n_\chi,n_f) >  \alpha_c(N) > \alpha_*(N-1,n_\chi,n_f).
\eeq
If this condition is satisfied, then, starting from a weakly coupled theory at
high energies, the coupling increases as the renormalization scale is lowered
until it becomes strong enough for chiral symmetry breaking to occur.  Below
the chiral symmetry breaking scale, the coupling is above  the fixed point of
the
unbroken gauge interactions, and as the renormalization scale is lowered
further,
the coupling decreases towards the new infrared fixed point.

As an example I will consider the model with
 $N=4$, $n_\chi=2$, and $n_f=0$. For this model
the estimate of the coupling
at the chiral symmetry breaking scale, $m$, is
\beq
\alpha(m) \approx \alpha_c={{2 \pi} \over {3 \,  C_2({\bf 10} ) }}
= {{4\pi}\over{27}} \approx  0.47 ~,
\label{crit}
\eeq
while $\alpha_*(4,2,0)\approx 0.61$, and $\alpha_*(3,2,0)\approx 0.42$.
(Another possible model is $N=4$, $n_\chi=1$,
and $n_f=8$, where $\alpha_*(4,1,8) \approx 0.58$, and
$\alpha_*(3,1,8) \approx 0.22$.) Above the scale $m$ there
are sixteen ${\bf 4}$'s and two
${\bf \overline{10}}$'s of  $SU(4)$. At the
chiral symmetry breaking scale $SU(4)$ breaks to $SU(3)$, and two
Dirac fermions get masses, leaving eighteen ${\bf 1}$'s,
fourteen ${\bf 3}$'s, and two
${\bf {\overline 6}}$'s of $SU(3)$.
Will any further condensation take place below this
scale?  According to the MAC hypothesis, the answer is no.  The
remaining condensation channels are less attractive (i.e. some of the
gauge bosons have become massive) than the channel that has already
condensed, and as the renormalization scale is moved towards the
infrared, the coupling grows weaker, not stronger.

If the estimate for $\alpha(m)$, equation (\ref{crit}), is correct, then  the
unbroken
gauge theory is not very strongly coupled.
Recall that by itself the long-range gauge force would lead
to an infinite number of bound states (labeled by $n$)
of the massive fermions with binding energies
given by:
\beq
E_n = - {{ m \,C_2({\bf  3} )^2 \alpha^2}\over {4 n^2}}= -{{4 m \alpha^2}\over
{9 n^2}} ~,
\eeq
where $\alpha=\alpha(m \alpha)$,
so the scalar (a $2P$ level) receives a long-distance contribution to
its splitting from the analogue of
the $\rho$ (a $1S$ level) equal to $m\alpha^2/3$.
The long-distance contribution to the splitting between
the analogue of the $\rho$ and the massless Goldstone bosons
(some of which are eaten by gauge bosons) comes solely from
hyperfine splitting, and is of order
$m\alpha^4$.  Thus the scalar is potentially as light as roughly $m/8$,
i.e. roughly forty times lighter than a naive scaling from QCD would suggest.

What about short-distance contributions to the scalar mass from
higher-dimension operators?  The implicit assumption of the above
discussion is that as the renormalization scale is lowered towards
the chiral symmetry breaking scale (and the coupling constant
grows), no irrelevant operators become relevant.  Whether or not
this happens can probably only be determined reliably in lattice
simulations.  The success of  non-relativistic quark models can be
taken as circumstantial evidence that short-distance contributions
(aside from instanton contributions)
to meson masses from higher-dimension operators are small in QCD.
Of course, in the example I have considered, the short distance interactions
are important for making the Goldstone bosons massless.

\section{Conclusions}
Using Hill and Marinelli's observation of the instanton contribution
to scalar masses in QCD, I have  pointed out that scalars
associated with chiral symmetry breaking will be lighter in
a class of chiral gauge theories than a naive scaling might
suggest.  I have also pointed out that in chiral gauge theories
that exhibit chiral symmetry breaking but not confinement,
the scalars will be even lighter. Finally a model was constructed
that is asymptotically free above the chiral symmetry breaking
scale, and (assuming the MAC hypothesis is correct) is
non-asymptotically free (hence not
confining) below this scale.
If the simulation of chiral gauge theories on the lattice becomes
feasible, then models like this may be useful in disentangling
chiral symmetry breaking and confinement.

On a more speculative note, model builders who try to extend the
standard model often end up with light scalars in their models.
The scalars are usually kept light by some form of fine-tuning.
The considerations above suggest that it may be possible to
construct models with light-composite scalars without resorting
to fine-tuning.

\noindent \medskip\centerline{\bf Acknowledgements}
I thank  S. Chivukula for inspiring the toy model and constructive criticism;
C. Hill for sending an early draft of his work; and finally A. Cohen, P.
Geiger,
K. Lane,  E. Swanson, and R. Sundrum for
helpful discussions.  I would also like to thank the Aspen Center
for Physics, where part of this work was completed.
This work was partially supported by the Department of Energy under
contract \#DE-FG02-91ER40676.

\end{document}